# SAI-BA-IoMT: Secure AI-Based Blockchain-Assisted Internet of Medical Things Tool to Moderate the Outbreak of COVID-19 Crisis

Mahender Kumar[1], Ruby Rani[2]

**Abstract:** Recently, an infectious disease, coronavirus disease 2019 (COVID-19), has been reported in Wuhan, China, and subsequently spread worldwide within a couple of months. On May 16, 2020, the COVID-19 pandemic has affected several countries in the world. In this article, we present how the amalgamation of blockchain, cryptography, internet of medical things (IoMT), and artificial intelligence (AI) technologies can address such an issue in the event of the COVID-19 pandemic. Further, we propose a secure AI-based blockchain-assisted IoMT (SAI-BA-IoMT) model for the healthcare system in the COVID-19 crisis. The paper also examines the post-corona crisis that the world could be experienced after the pandemic. Additionally, we exhibit the potential applications of the proposed model to resolve the difficulties originated from coronavirus.

## 1 Introduction

Nowadays, the entire world is confronting a critical health crisis due to a biological attack, COrono VIrus Disease (COVID-19) [1]. It is a kind of respiratory disease whose impact will seemingly be more distinguishable than that of the existing severe acute respiratory syndrome (SARS), held in 2003. Several countries are extrapolating the infected-control and public-health measures to control the pandemic. A large number of potentially infected people have been isolated (quarantined), and countries have been locked down to overcome the unexpected outbreak of coronavirus diseases. Although these approaches are partially practical since lockdown has affected the economy of many countries. Besides, it is well-known that the outbreak of COVID-19 in many countries such as Italy, the USA, and Span, has been brutally affected their medical instructions also, where the case of infected patients overwhelmed the hospitals [2].

Due to COVID-19, the world is currently suffering from severe problems such as the shortage of hospitals and medical staff, global economic loss, post-corona disease, and security and privacy of patient information. To tackle the problems as mentioned above, we are exploring four emerging potential technologies: blockchain [3], Internet of Medical Things (IoMT) [4], Artificial intelligence [5], [6], and lightweight cryptosystem [7]. Our focus is to employ such digital technologies, help the people in moderating the outbreak of pandemics. Amalgamating blockchain and artificial intelligence (AI) with healthcare covers a diverse area of healthcare services, in which wireless body area network (WBAN) is an essential part of the perspective of IoT application. Recently, Chang et al. [7] discussed some security attacks associated with COVID-19, such as a record of infected people, donation transparency, spreading false information, and social distancing, where blockchain can solve to address such security attacks. Here, we observed a smart monitoring system that can help peoples to reduce the outbreak of COVID-19 pandemic. Inspired by this idea, in this paper, we present a smart mounting system that integrates the blockchain, AI, lightweight cryptosystem, and Internet of Medical things to addresses the problem associated with COVID-19.

In the following, the background, including the brief of novel emerging technologies, will be discussed in Section 2. We will analyze the COVID-19 crisis and discuss the problems that currently world has been facing due to the COVID-19 in Section 3. Section 4 discusses the amalgamated technology that we have been proposed. The recent impact technologies in the current crisis and their solutions are explained in Section 5. The paper concludes in Section 6.

## 2 Background

Here, we discuss the recent emerging technologies that we are focusing in order to address the issue associated to the COVID-19

[1] Mahendjnu1989@gmail.com, School of Computer and System Sciences, Jawaharlal Nehru University, Delhi, 1100167, India
[2] Ruby73_scs@mail.jnu.ac.in, School of Computer and System Sciences, Jawaharlal Nehru University, Delhi, 1100167, India

*Internet of Medical Things*. The IoMT, in healthcare, connects several wearable or implanted sensors (have limited computational power and storage data) on the patient's body that continuously senses and sends data to the personal server (smartphone) [8]. The personal server collects data from each sensor and helps the WBAN for transmitting data to the medical server via the internet. After receiving data, the medical server processes and stores it and suggests an appropriate physician based on the severity of the patient. The medical server is a central database that may inevitably result in more significant damage if the system is exposed to a hacking attack. To address this issue, we adopt blockchain (a public ledger of transactions), provides a decentralized database of records, and share data among each participate.

*Blockchain*. Blockchain is a chain of blocks linked to the previous one using cryptographic hash and has a set of immutable transactions [9]. Each transaction is verified by the consensus protocol (Proof of Work (PoW), Proof of Stake (PoS), and Practical Byzantine Fault Tolerance (PBFT)) of a majority of the participants in the network. Once the data is stored, it can never be modified/deleted, and so easily tracked. One of the most famous examples in practice that uses blockchain is the cryptocurrency, Bitcoin [10]. Blockchain has a wide range of applications in several areas, including peer-to-peer insurance, online voting, smart contracts, and its adoption in the health care system is also being discussed. Since the sensor and medical server have different computational power and storage space, it is also challenging to ensure the security and privacy of data in such a lightweight environment.

*Public key cryptosystem*. Public key cryptosystem holds two keys: private (decryption/signature) key public (encryption/verification) key. A private key is kept secret to the user, and publically key is publically broadcast to the available receipt in the system. Both keys are mathematically linked in such a way that it is negligible to compute the private key on the given public. In such a cryptosystem, a private key is responsible for ensuring authenticity, non-repudiation, and authorization, while the public key ensures the confidentiality and integrity of data. There have discussed several examples of public-key cryptosystems, such as RSA, ElGamal, and identity-based crypto stem (IBC) that are implemented on the distinct mathematical tool, such as modular exponentiation, elliptic curve, and bilinear pairing [11]. In the proposed technology, we are adopting IBC implemented on the elliptic curve as a cryptosystem to secure transactions during communication. It is considered as the lightweight cryptosystem for IoT [12].

*Artificial Intelligence*. Besides, we believed that detecting diseases at an early stage can enable us to overcome and treat them appropriately. In this direction, we introduced a Diagnosis Expert System (DES) instead of a medical server [13], [5]. In the proposed work, the diagnosis expert system is an artificial intelligence-based computer system that has a decision-making ability as similar to a human expert, which fetches the data from the blockchain and identifies the treatment accurately depends on the method that is used in diagnosing the diseases.

We are amalgamating these emerging technologies to present a state-of-the-art smart, automated monitoring healthcare system, in which IoMT is responsible for providing a network to collect patient data from the patient and recommending a prescription suggested by diagnosis expert system. We will adopt a lightweight cryptosystem to ensure the integrity, confidentiality, and authenticity of medical data during each transaction.

## 3 Outbreak of COVID-19: Analysis

In late 2019, the first case of the corona has been seen in Wuhan, China. It had been circulating within the population since January and has been spread globally at an exponential rate. On March 11, 2020, COVID-19 had affected around 118,000 people in 114 countries, and 4291 people have lost their lives. Accordingly, WHO characterized COVID-19 as a pandemic [14]. After China, Italy became the second hotspot of most significant infected cases COVID-19, with a very high fatality rate of 7.2% (1625 deaths/22512 cases), which was substantially higher on comparing in China (2.3%) on March 17, 2020 [15].

On March 27, 2020, the USA was the first to reach the 100,000 cases, which is double in the next five days and reach 1 million with a 5.8% fatality rate in the next 31 days [16]. As of May 16, 2020, coronavirus disease, 2019 (COVID-19), has been infected to around 4.3 million people worldwide, carrying a fatality rate of around 7.04%, as shown in Table 1 and Fig. 1. Currently, the US has the highest number of infected cases (1,522,149), which is 31.85% of total cases worldwide, with a 5.96% fatality rate, shown in Fig. 2.

Fig. 3 and Table 1 show that the period from March 12-April 3, 2020, was the transition period where the newly infected cases were linearly increasing. There can be seen a constant growth of newly infected cases (around 800,000) from April 13, 2020. The fatality rate has been increased from 1.67% to 7.18% since April 27, 2020. There is a 0.24% decline of fatality rate from April 27, 2020, to May 16, 2020, as illustrated in Table 1. Fig. 4 and Table 1 show that the recovered patients' cured rate, where we can see that graph is fortunately linearly increasing.

Table 1. Week wise summary of infected case due to COVID-19 around the world till 16 May 2020

| Date | Total cases | New Cases per week | Total deaths | New Deaths per week | Total cases PM$ | New cases PM | Total deaths PM | New deaths PM | Total Fatality rate# |
|---|---|---|---|---|---|---|---|---|---|
| 2019-12-31 | 27 | 27 | 0 | 0 | 0.003 | 0.003 | 0 | 0 | 0 |
| 2020-01-07 | 59 | 32 | 0 | 0 | 0.008 | 0 | 0 | 0 | 0 |
| 2020-01-14 | 60 | 1 | 1 | 1 | 0.008 | 0 | 0 | 0 | 1.67 |
| 2020-01-21 | 392 | 332 | 6 | 5 | 0.05 | 0.02 | 0.001 | 0 | 1.53 |
| 2020-01-28 | 4587 | 4195 | 106 | 100 | 0.588 | 0.227 | 0.014 | 0.003 | 2.31 |
| 2020-02-05 | 24522 | 19935 | 493 | 387 | 3.146 | 0.501 | 0.063 | 0.008 | 2.01 |
| 2020-02-12 | 45177 | 20655 | 1115 | 622 | 5.796 | 0.266 | 0.143 | 0.012 | 2.47 |
| 2020-02-19 | 75191 | 30014 | 2012 | 897 | 9.646 | 0.239 | 0.258 | 0.018 | 2.68 |
| 2020-02-26 | 80995 | 5804 | 2762 | 750 | 10.391 | 0.111 | 0.354 | 0.008 | 3.41 |
| 2020-03-03 | 90865 | 9870 | 3118 | 356 | 11.657 | 0.231 | 0.4 | 0.009 | 3.43 |
| 2020-03-10 | 114235 | 23370 | 4021 | 903 | 14.655 | 0.582 | 0.516 | 0.027 | 3.52 |
| 2020-03-17 | 180094 | 65859 | 7112 | 3091 | 23.104 | 1.626 | 0.912 | 0.077 | 3.95 |
| 2020-03-24 | 377968 | 197874 | 16387 | 9275 | 48.49 | 5.097 | 2.102 | 0.228 | 4.34 |
| 2020-03-31 | 777187 | 399219 | 37911 | 21524 | 99.706 | 7.93 | 4.864 | 0.5 | 4.88 |
| 2020-04-06 | 1245601 | 468414 | 69859 | 31948 | 159.799 | 9.146 | 8.962 | 0.6 | 5.61 |
| 2020-04-13 | 1807256 | 561655 | 115182 | 45323 | 231.854 | 9.287 | 14.777 | 0.676 | 6.37 |
| 2020-04-20 | 2350993 | 543737 | 167483 | 52301 | 301.61 | 9.302 | 21.487 | 0.646 | 7.12 |
| 2020-04-27 | 2915977 | 564984 | 209254 | 41771 | 374.093 | 10.677 | 26.845 | 0.504 | 7.18 |
| 2020-05-04 | 3468047 | 552070 | 247163 | 37909 | 444.918 | 10.082 | 31.709 | 0.473 | 7.13 |
| 2020-05-09 | 3968658 | 5000611 | 274290 | 27127 | 500.161 | 11.469 | 35.189 | 0.647 | 7.04 |
| 2020-05-16 | 4495412 | 526754 | 312120 | 37830 | 521.880 | 11.845 | 37.521 | 0.592 | 6.94 |

$per million
#Total number of death from COVID-19

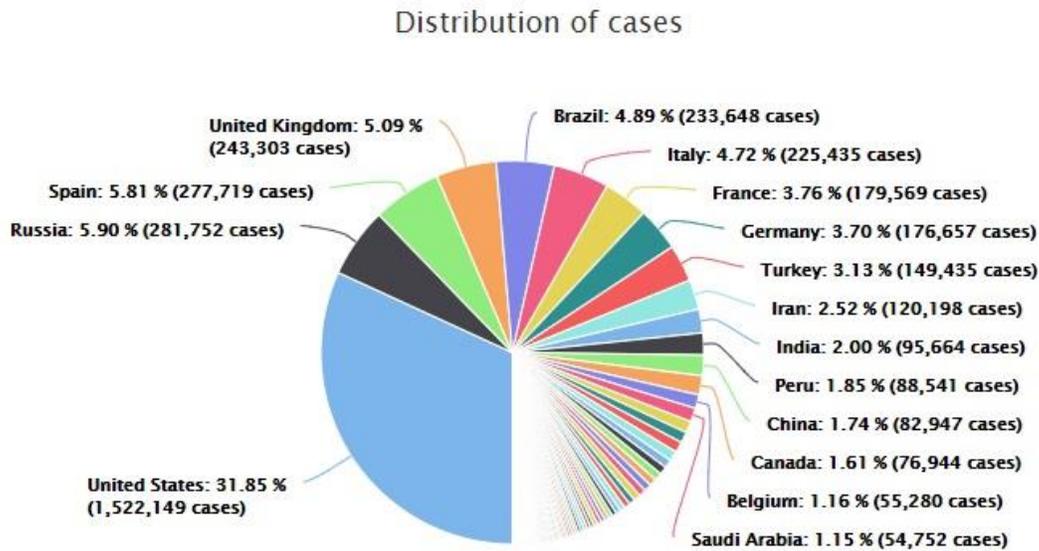

Fig. 2 Infected cases distribution on 16, May 2020 (source: https://www.worldometers.info/coronavirus) [16]

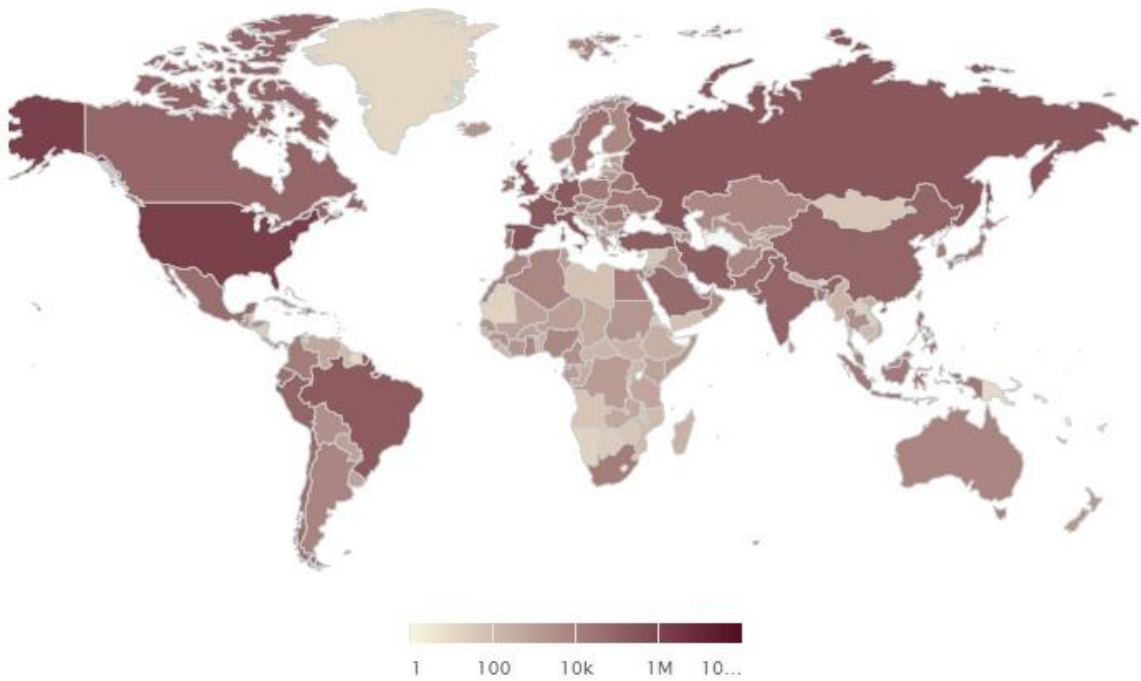

Fig. 1 Total Number of cases world wise on 16, May 2020 (source: https://www.worldometers.info/coronavirus) [16]

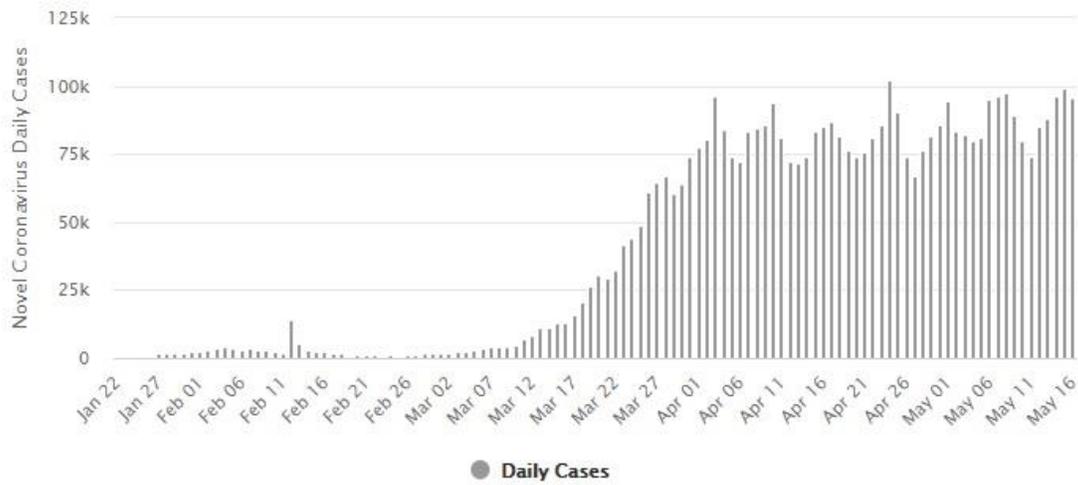

Fig. 3. Daily world wise infected cases on 16, May 2020 (source: https://www.worldometers.info/coronavirus) [16]

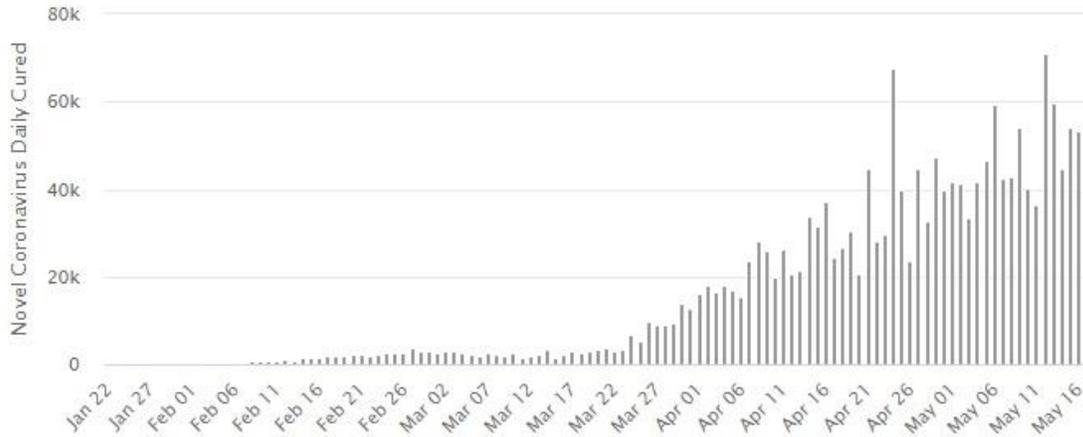

Fig. 4. Daily world wise recovered on 16, May 2020 (source: https://www.worldometers.info/coronavirus) [16]

Over four months into 2020, the world is encountering an extremely health crisis: the outbreak of a respiratory disease-coronavirus (COVID-19). The impact of COVID-19 will probably be higher than that of the severe acute respiratory syndrome (SARS) in 2003. Many countries tackled this new kind of crisis by extrapolating the traditional infected-control and public-health measures to control the COVID-19 pandemic. It ranges from extreme quarantine measure (e.g., in China and India) to painstakingly detailed contact tracing with thousands of contact tracers (e.g., South Korea and Singapore) [1]. Currently, a large number of potentially infected people need to be isolated (quarantined) in order to overcome the unexpected outbreak of coronavirus diseases [2]. Such approaches could not be practical to moderate the COVID-19 outbreak while the economy of many countries has been affected due to lockdown. Could a new digital technology be helped the system for COVID-19? There should be an urgent need for an effective way. In this direction, we explore four potential applications of digital technologies in order to augment a strategy for tackling COVID-19: blockchain, AI, public-key cryptosystem, and the Internet of Medical Things. Now, we present some problems related to the COVID-19 crisis.

1. *Post-corona diseases*. Recently, researchers have uncovered the post-corona situation that will be worst for the patient who has been once infected from viruses. The people who recovered from the virus are more susceptible to diseases, like Heart attack, High BP, respiratory disease, and diabetes.
2. *Lack of medical resources.* Secondly, many countries have been lacking in medical equipment, hospitals, and medical staff due to the outbreak of COVID-19.
3. *Destructed global economic*. However, they locked down the whole country to overcome this difficult situation, but it came with the countries' financial loss. Thus, the third issue associated with COVID-19 is the destruction of the global economy.
4. *Privacy of Patient' information*. Each participant, including patient and hospital, is concerned about the privacy of patient information, including identity, health data, suggested physician, type of treatment.

## 4 Proposed model for COVID-19 crisis

Here, we present the architecture of the proposed secure AI-based Blockchain-assisted IoMT (SAI-BA-IoMT) model for health care system in COVID-19, as shown in Fig. 5. It consists of six entities: user, sensor, personal digital assistant (PDA), blockchain, permissioned agent, medical server, and diagnose the expert system.

A *user* can be a patient who is suffering from a virus or a person who has recovered from a virus. He implants or wears the set of *sensors* in his body in order to collect the medical data. Along with this, the user also has a *personal digital assistant* (PDA), such as a smartphone that collects the medical data from each sensor and stores in its storage. User can visually monitor his health status from PDA, encrypts it and periodically uploads it in *blockchain* in the form of block via the internet. Here, we consider each communication as a transaction that includes the medical data, timestamp, and information of the previous block.

There are two kinds of *permissioned agents*: validating agent and recording agent. The *validating agents* (VA) are the group of nodes responsible for validating the transaction. The only transaction validated through the majority of validating agents is eligible to store in the blockchain. After validating, the recording agents store the data in the form of block and allows only those participating in the network to access who has permission.

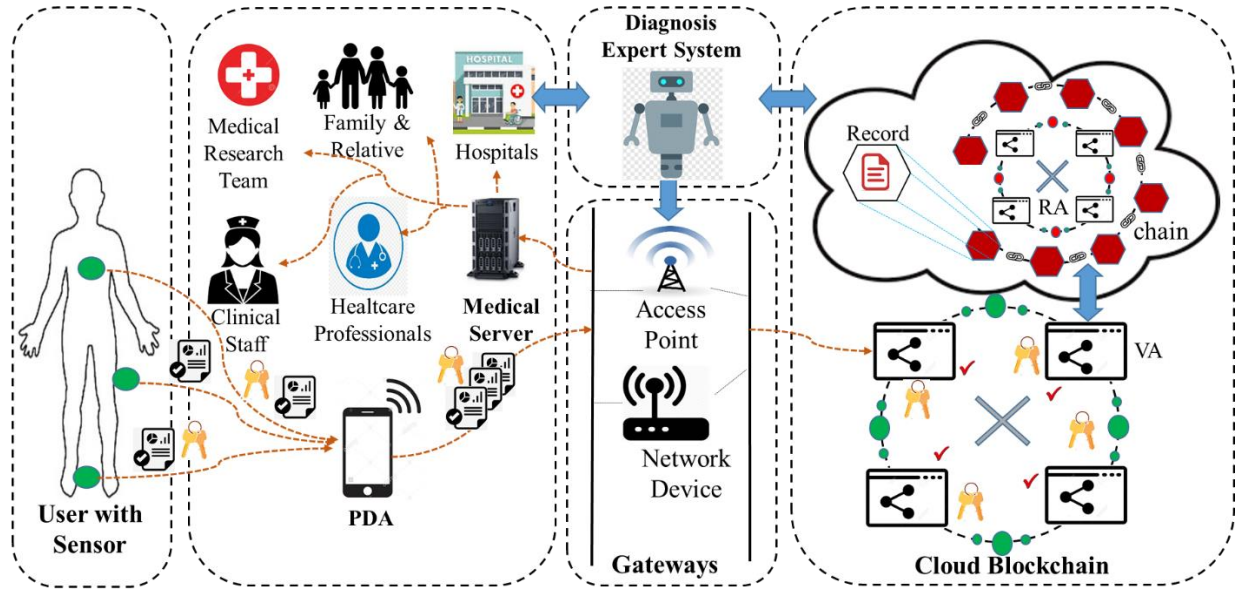

Figure 5. Architecture of proposed secure AI-based blockchain-enabled-IoMT model

In the proposed model, each participant, including medical staff, hospital, diagnosis expert system, manage a ledger for each block in the network. A *diagnosis expert system* is an AI-based computer system with a decision-making ability similar to a human expert. The DES fetches the data from the blockchain, decrypts it, and processes it to identify the type of diseases. It also suggests the prescription to a patient, physician if necessary, or command the sensors on the patient's body remotely, if the patient is in severe condition. This completes the brief of our proposed model.

# 5 Advantages of proposed SAI-BA-IoMT model in COVID-19

Here, we discuss how our proposed model can assist people, medical institution and country to handle the event of COVID-19 crisis.

## 5.1 Lack of medical resources

Many countries have been brutally affected by the COVID-19 outbreak. Due to a lack of hospitals, medical equipment, and medical professionals. They have unable to diagnose every infected patient in their country, so many people have died due to a lack of hospitals and doctors. The proposed model can resolve such a crisis by monitoring and diagnosing hundreds and thousands of patients simultaneously. IoMT technology is responsibility is to collect data between patients and hospitals. Blockchain manages the patient's medical record and provides a distributed database. The public key cryptosystem ensures the integrity, confidentiality, and authenticity of data. AI is responsible for suggesting the type of physician, predicting kind of disease, and recommending the prescription based on the patient's medical record.

## 5.2 Decentralized database

The proposed SAI-BA-IoMT model adopts blockchain technology to eliminate the central database. In the existing medical system, the medical report needs to be going through many levels to reach the hospital, which takes a couple of days to verify and treatment. On the other side, the proposed model's database is distributed and accessible to those authorized participants who have permission to manage the ledger. There is no need for such a level of validation, and hence, the response time to diagnose a patient became significantly efficient. Secondly, since the database is distributed, there is less risk of data loss due to a single point failure attack. Third, since each participant has ledger in its storage, the proposed model does not require a central database to be online all time.

## 5.3 Privacy of Patient information

The proposed model uses a lightweight public-key cryptosystem, especially an identity-based cryptosystem implemented on the elliptic curve cryptosystem, to preserve the privacy of patient information. The IBC does not require to verify the recipient's public key. The ECC-based arithmetical operations are around 20 times faster, in terms of computational overhead than modular exponentiation. Also, the 128-bit ECC key has the same security level as a 1024-bit RSA key. The integration of unique features of IBC and ECC makes them suitable for IoT applications.

## 5.4 Share economy

According to the department of economic and social affairs, the UN, the global supply chain, and international trade have been disrupted. The World Bank and credit-rating agencies have downgraded. As per the International Monetary Fund (IMF), the global economy is expected to down by over 3 percent in 2020. Due to COVID-19, the global economy could lose worth up to $8.8 trillion. In this respect, our proposed SAI-BA-IoMT model can be exploited to create decentralized, shared economy application that enables people to monetize their product (service) to create more wealth [17]. To address such problem, IoMT will acts as IoT to connect various products, blockchain managed the data, and AI-based medical expert system acts as an expert system.

*5.5 Inter-organization interaction*

Private blockchain, especially Hyperledger Fabric, allows transactions from one blockchain network to other networks. In our proposed model, we can enhance the blockchain with a private blockchain, which can allow interaction between two organizations in different environments. Using this model, international and national supply chain management has been implemented. This can improve economic conditions also.

*5.6 Post-corona advantages*

The proposed model enables the medical system to monitor the patient's health situation regularly who have infected from viruses previously. If needed, the proposed system diagnose them.

# 6 Conclusion

This paper analyzed the impact of new infectious disease, COVID-19, on the entire world. We observed that many countries, such as the USA, Italy, and China, have been brutally affected through such pandemic. In this paper, we have addressed the four major problems associated with COVID-19: the deficiency of medical hospitals and staff, post-corona disorders, global economy, and privacy of patient information. We have discussed how the amalgamation of blockchain, IoMT, and artificial intelligence technologies can address such an issue in the event of the COVID-19 pandemic. Further, we show how the suggested technology can defeat the problems associated with the disease.

**Funding.** This study was funded by the counsel of scientific and industrial research, and EMR-II of Jawaharlal Nehru University.

**Compliance with Ethical Standards**

This article does not contain any studies with human participants performed by any of the authors.

**Conflict of Interest**. Mahender Kumar declares that he has no conflict of Interest.
Ruby Rani declares that she has no conflict of interest.